\documentclass[11pt]{amsart}

\usepackage{amscd}
\usepackage{amsmath}
\usepackage{graphicx}
\usepackage{amsfonts}
\usepackage{amssymb}
\begin{document}

\title[maps preserving separability]
{Linear maps preserving separability of pure states}

\author{Jinchuan Hou}
\address[Jinchuan Hou]{Department of
Mathematics\\
Taiyuan University of Technology\\
 Taiyuan 030024,
  P. R. of China}
\email{jinchuanhou@yahoo.com.cn}
\author{Xiaofei Qi}
\address[Xiaofei Qi]{
Department of Mathematics, Shanxi University, Taiyuan 030006, P. R.
of China;} \email{xiaofeiqisxu@yahoo.com.cn}


\thanks{{\it 2010 Mathematics Subject Classification.} 47B49; 47N50; 47A80 }
\thanks{{\it Key words and phrases.}
Linear maps, affine maps, quantum states, separability,  tensor
products}
\thanks{This work is supported by National Natural Science Foundation
of China (11171249, 11101250) and  Youth Foundation of  Shanxi
Province (2012021004).}

\begin{abstract}
Linear maps preserving pure states of a quantum system of any
dimension are characterized. This is then used to establish a
structure theorem for linear maps that preserve separable pure
states in multipartite systems. As an application, a
characterization of separable pure state preserving affine maps is
obtained.

\end{abstract}
\maketitle

\section{Introduction}

A quantum state $\rho$ is a density operator acting on a complex
Hilbert space which is positive semidefinite and has trace 1.
Furthermore, $\rho$ is a pure state if $\rho^2=\rho$ (i.e., $\rho$
is a rank-1 projection); $\rho$ is a mixed state if
$\rho^2\not=\rho$. Denote by ${\mathcal S}(H)$ the set of all states
on a Hilbert space $H$. In quantum information theory we deal, in
general, with multipartite systems. The underlying space $H$ of a
multipartite composite quantum system is a tensor product of
underlying spaces $H_{i}$ of its subsystems, that is,
$H=H_{1}\otimes H_{2}\otimes\cdots \otimes H_{n}$. In the case $n=2$
the system is called a bipartite system. If $H$ and $K$ are finite
dimensional Hilbert spaces, $\rho\in{\mathcal S}(H\otimes K)$ is
said to be separable if $\rho$ can be written as
$$\rho=\sum_{i=1}^k p_i \rho_i\otimes \sigma _i, \eqno(0.1)$$
where $\rho_i$ and $\sigma_i$ are states on $H$ and $K$
respectively, and $p_i$ are positive numbers with $\sum
_{i=1}^kp_i=1$. Otherwise, $\rho$ is said to be inseparable or
entangled (ref. \cite{BZ, NC}). For the case that at least one of
$H$ and $K$ is of infinite dimension,  by  Werner \cite{W},  a state
$\rho$ acting on $H\otimes K$ is called separable if it can be
approximated in the trace norm by  states of the form (0.1).
Otherwise, $\rho$ is called an entangled state. The full
separability of multipartite states can be defined similarly.

Entanglement is a basic physical resource to realize various quantum
information  and quantum communication tasks \cite{Hor1, Hor2, NC}.
So it is important to determine whether or not a state in a
composite system is separable, which is also a very difficult task
in this field. Thus, this makes it interesting to find linear maps
sending states to states, which will simplify a given state so that
it is easier to detect the entanglement in it. Clearly, such linear
maps  should  leave the separability of states invariant. So, this
proposes the question of studying linear preservers of separable
states. This question was attacked in \cite{FLPS} for the finite
dimensional systems. Let ${\bf H}_{N}$ be the real linear space of
all $N\times N$ Hermitian matrices. It was shown in \cite{FLPS}
that, if a {\it surjective} linear map $\Phi: {\bf
H}_{n_1n_2}\rightarrow{\bf H}_{n_1n_2}$ preserves separable pure
states  in the bipartite system ${\mathbb C}^{n_1}\otimes {\mathbb
C}^{n_2}$, then $\Phi$ sends product states to product states, that
is, $\Phi(A_1\otimes A_2)=\psi_1(A_{p_1})\otimes \psi_2(A_{p_2})$,
where $(p_1,p_2)$ is a permutation of $(1,2)$, $n_j=n_{{p_j}}$ and
$\psi_j: M_{n_j}\rightarrow M_{n_j}$ is a linear map of the form
$X\mapsto U_jXU_j^*$ or $X\mapsto U_jX^{\rm t}U_j^*$ for a unitary
matrix $U_j\in M_{n_j}$. Here $X^{\rm t}$ denotes the transposed
matrix of $X$. A similar result holds for finite dimensional
multipartite systems.

The purpose of the present paper is to characterize general linear
maps that preserve separable pure states   for both finite and
infinite dimensional systems. We remark that the results for finite
dimensional systems are somewhat different from that for infinite
dimensional systems   because the linear maps on infinite
dimensional spaces may not be continuous.

Let ${\mathcal T}(H)$ be the Banach space of trace-class operators
on a complex Hilbert space $H$ endowed with the trace-norm
$\|\cdot\|_{\rm Tr}$. Denote by ${\mathcal T}_{\rm sa}(H)$ and
${\mathcal F}_{\rm sa}(H)$ the subspace of self-adjoint operators
and finite-rank self-adjoint operators in ${\mathcal T}(H)$,
respectively. Denote by ${\mathcal Pur}(H)$ the set of pure states
(i.e., rank-one projections) on $H$. We first consider in Section 2
the question of characterizing linear maps preserving pure states
since this is basic for the study of our main question. Let $H$ and
$K$ be Hilbert spaces of any dimension. It is shown that a linear
map $\Phi:{\mathcal T}_{\rm sa}(H)\rightarrow{\mathcal T}_{\rm
sa}(K)$ satisfies $\Phi({\mathcal Pur}(H))\subseteq {\mathcal
Pur}(K)$ if and only if $\Phi$ either has the form $A\mapsto{\rm
Tr}(A)R+\phi(A)$ for all $A\in{\mathcal T}_{\rm sa}(H)$; or $\dim
H\leq\dim K$ and $\Phi$ has the form $A\mapsto UAU^*+\phi(A)$ for
all $A\in{\mathcal T}_{\rm sa}(H)$, where $\phi: {\mathcal T}_{\rm
sa}(H)\rightarrow{\mathcal T}_{\rm sa}(K)$ is a linear map vanishing
on each finite-rank operator, $R\in{\mathcal Pur}(K)$ and
$U:H\rightarrow K$ is a linear or conjugate linear isometry (Theorem
2.2).  Particularly, if $\Phi$ preserves pure states in both
directions, then $\Phi$ has  the second  form  with $U$ unitary or
conjugate unitary. The main result of this section generalizes a
result in \cite{FLPS} for finite dimensional case.

In Section 3 we discuss the question of characterizing linear maps
preserving separable pure states of bipartite systems. Let
${\mathcal S}_{\rm sep}(H\otimes K)$ stand for the convex set of all
separable states on $H\otimes K$. Denote by ${\mathcal T}_{\rm
sep}(H\otimes K)$ the linear manifold generated by ${\mathcal
S}_{\rm sep}(H\otimes K)$ and ${\mathcal F}_{\rm sep}(H\otimes
K)={\mathcal T}_{\rm sep}(H\otimes K)\cap {\mathcal F}_{\rm
sa}(H\otimes K)$. It is obvious that ${\mathcal T}_{\rm
sep}(H\otimes K)\subseteq {\mathcal T}_{\rm sa}(H\otimes K)$. Our
main result gives a characterization of linear maps from ${\mathcal
T}_{\rm sa}(H\otimes K)$ into itself which preserve separable pure
states ({\it not necessarily in both directions}). It turns out such
maps have one of nine forms on ${\mathcal F}_{\rm sep}(H\otimes K)$
(see Theorem 3.2). However, for most situations they have a standard
form ((6) or (7) in Theorem 3.2). As an application, we get a
characterization of affine maps between convex sets of states which
preserve separable pure states in both directions. Such a map is
either of the form $\Phi(\rho)=(U_1\otimes
U_2)\Lambda(\rho)(U_1\otimes U_2)^*$ for all $\rho\in{\mathcal
S}_{\rm sep}(H\otimes K)$
 or  of the form  $\Phi(\rho)=(U_1\otimes U_2)\Lambda(\theta (\rho))(U_1\otimes
 U_2)^*$ for all $\rho\in{\mathcal
S}_{\rm sep}(H\otimes K)$, where $U_1$ and $U_2$ are  unitary
operators on $H$ and $K$, respectively, $\Lambda $ is one of the
identity map, the transpose, the partial transpose with respect to
any fixed product orthonormal basis of $H\otimes K$, and  $\theta:
{\mathcal T}(H\otimes K)\rightarrow{\mathcal T}(K\otimes H)$ is the
swap determined by $\theta(A\otimes B)=B\otimes A$.

In Section 4, a brief discussion of the question for multipartite
systems is given. Some results similar to those in bipartite systems
in Section 3 are presented.

\section{Linear maps preserving pure states}

The main purpose of this section is to characterize linear
preservers of pure states for infinite dimensional systems, which
are also needed to characterize separable pure state preservers in
the next section.

The following proposition comes from \cite{FLPS}, which can be
viewed as a characterization of linear preservers of pure states for
finite dimensional systems.

 Let ${\bf H}_m$ be the real linear space of all
$m\times m$ Hermitian matrices and let ${\mathcal P}_m$ be the set
of all rank-1 $m\times m$ projection matrices.

{\bf Proposition 2.1.} {\it Suppose that $\phi :{\bf H}_m\rightarrow
{\bf H}_n$ is linear and satisfies $\phi({\mathcal
P}_m)\subseteq{\mathcal P}_n$. Then one of the following holds:}

(i) {\it There is $Q\in{\mathcal P}_n$ such that $\phi(A)={\rm
Tr}(A)Q$ for all $A\in{\bf H}_m$.}

(ii) {\it $m\leq n$ and there is a matrix $U\in M_{n\times m}$ with
$U^*U=I_m$ such that $\phi(A)=UAU^*$ for all $A\in{\bf H}_m$, or
$\phi(A)=UA^{\rm t}U^*$ for all $A\in{\bf H}_m$.}

By using a result due to \cite{L}, we can generalize Proposition 2.1
to the infinite dimensional case. Recall that a linear map
$V:H\rightarrow K$ is an isometry if $\|Vx\|=\|x\|$ for all $x\in
H$, or equivalently, $V^*V=I_H$, the identity operator on $H$. A
conjugate linear isometry is defined similarly.

The following is the main result of this section.

{\bf Theorem 2.2.} {\it Let $H$ and $K$ be    Hilbert spaces of any
dimension. Suppose that $\Phi:{\mathcal T}_{\rm
sa}(H)\rightarrow{\mathcal T}_{\rm sa}(K)$ is a real linear map.
Then $\Phi$ satisfies $\Phi({\mathcal Pur}(H))\subseteq {\mathcal
Pur}(K)$ if and only if there exists a linear map $\phi: {\mathcal
T}_{\rm sa}(H)\rightarrow{\mathcal T}_{\rm sa}(K)$ vanishing on each
finite-rank operator and one of the following holds:}

(i) {\it There is some $R\in{\mathcal Pur}(K)$   such that
$\Phi(A)={\rm Tr}(A)R+\phi(A)$ for all $A\in{\mathcal T}_{\rm
sa}(H)$.}

(ii) {\it $\dim H\leq \dim K$ and there is a linear or conjugate
linear isometry $U:H\rightarrow K$ such that $\Phi(A)=UAU^*+\phi(A)$
for all $A\in{\mathcal T}_{\rm sa}(H)$.}

{\bf Proof.} Only the ``only if" part should be checked.

Suppose  that $\Phi$ preserves pure states. By Proposition 2.1, we
may assume that $\dim H=\infty$. Define a map $\Psi:{\mathcal
T}_{\rm sa}(H\oplus K)\rightarrow{\mathcal T}_{\rm sa}(H\oplus K)$
given by
$$\Psi(S)=\Psi(\begin{bmatrix}A&C\\
C^*&B
\end{bmatrix})=\begin{bmatrix}0&0\\0&\Phi(A)\end{bmatrix}$$
for all
$$S=\begin{bmatrix}A&C\\
C^*&B
\end{bmatrix}\in {\mathcal T}_{\rm sa}(H\oplus K)\ {\rm with}\ A\in{\mathcal T}_{\rm sa}(H).$$
It is obvious that $\Psi$ is linear and $\Psi(A\oplus 0)=0\oplus
\Phi(A)$ for all $A\in{\mathcal T}_{\rm sa}(H)$. Moreover, $\Psi$ is
rank one decreasing, that is, ${\rm rank}(\Psi(S))\leq1$ whenever
${\rm rank}(S)=1$, since $\Phi({\mathcal Pur}(H))\subseteq {\mathcal
Pur}(K)$. It follows that $\Psi({\mathcal F}_{\rm sa}(H\oplus
K))\subseteq{\mathcal F}_{\rm sa}(H\oplus K)$, where ${\mathcal
F}_{\rm sa}(H)$ stands for the set of all finite-rank self-adjoint
operators on $H$. Hence, by \cite[Theorem 2.10]{L}, we get that one
of the following is true:

(1) $\Psi(S)=f(S)Q$ for all $S\in{\mathcal F}_{\rm sa}(H\oplus K)$,
where $f:{\mathcal F}_{\rm sa}(H\oplus K)\rightarrow{\mathbb R}$ is
a linear map and $Q$ is a rank one projection;

(2) $\Psi(x\otimes x)=\lambda Tx\otimes Tx$ for all $x\in H\oplus
K$, where $\lambda$ is a nonzero real number and $T:H\oplus
K\rightarrow H\oplus K$ is a linear or conjugate linear operator.

If (1) holds, then there exists some $R\in{\mathcal Pur}(K)$ such
that $Q=0\oplus R$ and $\Phi(A)=g(A)R$ for all $A\in{\mathcal
F}_{\rm sa}(H)$, where $g(A)=f(A\oplus 0)$. Since $\Phi({\mathcal
Pur}(H))\subseteq {\mathcal Pur}(K)$, we have $g(P)=1$ for all
$P\in{\mathcal Pur}(H)$. Now, for any $A\in{\mathcal F}_{\rm
sa}(H)$, write $A=\sum_{i=1}^n \lambda_i P_i$, which is the spectral
decomposition of $A$. It follows from the linearity of $\Phi$ that
$g(A)=g(\sum_{i=1}^n \lambda_i P_i)=\sum_{i=1}^n \lambda_i
g(P_i)=\sum_{i=1}^n \lambda_i={\rm Tr}(A)$, that is, (i) holds for
all finite-rank operators. Next let us show that the statement (i)
of Theorem 2.2 holds for all $A\in{\mathcal T}_{\rm sa}(H)$. Note
that $\Phi$ is bounded on the normed space ${\mathcal F}_{\rm
sa}(H)$ endowed with the trace-norm. In fact, $\|\Phi|_{{\mathcal
F}_{\rm sa}(H)}\|\leq \|R\|$. Let $\widehat{\Phi}:{\mathcal T}_{\rm
sa}(H)\rightarrow{\mathcal T}_{\rm sa}(K)$ be the bounded linear map
defined by $\widehat{\Phi}(A)={\rm Tr}(A)R$ and let
$\phi=\Phi-\widehat{\Phi}$. Then, $\phi(F)=0$ for each
$F\in{\mathcal F}_{\rm sa}(H)$ and  $\Phi(A)={\rm Tr}(A)R+\phi(A)$
for all $A$, as desired.

Now assume that (2) holds. Note that $\Psi(A\oplus 0)=0\oplus
\Phi(A)$. Then $\Phi$ has the form $\Phi(x\otimes x)=\lambda
Ux\otimes Ux$ for all $x\in H$, where $U$ is the part of $T$
restricted to $H$. Next we  prove that $U$ is bounded. In fact,
since $\Phi({\mathcal Pur}(H))\subseteq{\mathcal Pur}(K)$, we have
$$\lambda\langle
Ux,Ux\rangle=\lambda\|Ux\|^2=1\eqno(2.1)$$ for all unit vectors
$x\in H$, which implies $\lambda>0$. Without loss of generality, we
may assume that $\lambda=1$. Since $U$ is linear or conjugate
linear, by Eq.(2.1), we get $\|Ux\|=\|x\|$ for all $x\in H$.  It
follows that $U$ is bounded and $U^*U=I_H$, that is, $U$ is an
isometry or conjugate isometry. Then $\Phi(x\otimes x)=U(x\otimes
x)U^*$, and consequently, $\Phi(A)=UAU^*$ for all $A\in{\mathcal
F}_{\rm sa}(H)$. Let $\phi: {\mathcal T}_{\rm sa}(H)\rightarrow
{\mathcal T}_{\rm sa}(H)$ be the linear map defined by
$\phi(A)=\Phi(A)-UAU^*$ for every $A$. Then $\Phi(A)=UAU^*+\phi(A)$
for all $A\in{\mathcal T}_{\rm sa}(H)$, that is, (ii) of Theorem 2.2
holds.

The proof of the theorem is complete.\hfill$\Box$

{\bf Remark 2.3.} If the linear map $\phi:{\mathcal T}_{\rm
sa}(H)\rightarrow {\mathcal T}_{\rm sa}(H)$ is not zero, then both
$\Phi$ and $\phi$ are not continuous because ${\mathcal F}_{\rm
sa}(H)$ is dense in ${\mathcal T}_{\rm sa}(H)$. Such linear maps
exist. For example, take any nonzero linear map
$\widehat{\phi}:{\mathcal T}_{\rm sa}(H)/{\mathcal F}_{\rm sa}(H)
\rightarrow{\mathcal T}_{\rm sa}(K)$, and let $\phi$ be defined by
$\phi(A)=\widehat{{\phi}}(\pi (A))$ for any $A\in{\mathcal T}_{\rm
sa}(H)$, where $\pi :{\mathcal T}_{\rm sa}(H)\rightarrow {\mathcal
T}_{\rm sa}(H)/{\mathcal F}_{\rm sa}(H) $ is the quotient map.

The following corollary is immediate from Theorem 2.2.

{\bf Corollary 2.4.} {\it Let $H$ and $K$ be    Hilbert spaces of
any dimension. Suppose that $\Phi:{\mathcal T}_{\rm
sa}(H)\rightarrow{\mathcal T}_{\rm sa}(K)$ is a bounded real linear
map. Then $\Phi$ satisfies $\Phi({\mathcal Pur}(H))\subseteq
{\mathcal Pur}(K)$ if and only if  one of the following holds:}

(i) {\it There is some $R\in{\mathcal Pur}(K)$   such that
$\Phi(A)={\rm Tr}(A)R$ for all $A\in{\mathcal T}_{\rm sa}(H)$.}

(ii) {\it There is a linear or conjugate linear isometry
$U:H\rightarrow K$ with $U^*U=I_H$ such that $\Phi(A)=UAU^*$ for all
$A\in{\mathcal T}_{\rm sa}(H)$.}

 Note that a bijective affine map from ${\mathcal S}(H)$ onto
 ${\mathcal S}(K)$
preserves pure states in both directions. So the following corollary
is a generalization of Kadison's characterization of affine
isomorphisms on ${\mathcal S}(H)$, which says that a bijective
affine map has the form $\rho\mapsto U\rho U^*$, where $U$ is a
unitary or conjugate unitary operator (See, for instance,
\cite[Theorem 8.1]{BZ}).

{\bf Corollary 2.5.} {\it Let $H$ and $K$ be    Hilbert spaces of
any dimension. Suppose that $\Phi:{\mathcal
S}(H)\rightarrow{\mathcal S} (K)$ is an affine map. Then $\Phi$
satisfies $\Phi({\mathcal Pur}(H))\subseteq {\mathcal Pur}(K)$ if
and only if one of the following holds:}

(i) {\it There is some $R\in{\mathcal Pur}(K)$   such that
$\Phi(\rho)= R$ for all $\rho\in{\mathcal S}(H)$.}

(ii) {\it There is a linear or conjugate linear isometry
$U:H\rightarrow K$  such that $\Phi(\rho )=U\rho U^*$ for all
$\rho\in{\mathcal S}(H)$.}

{\bf Proof.} We need only  show that if $\Phi:{\mathcal
S}(H)\rightarrow{\mathcal S} (K)$ is   affine and preserves pure
states, then $\Phi$ has the form (i) or the form (ii) stated in the
corollary.

To do this, note that the affinity of $\Phi$ allow us to extend it
to a linear map (see Ref. \cite{FLPS} for details), still denoted by
$\Phi$, from $ {\mathcal T}_{\rm sa}(H)$ into ${\mathcal T}_{\rm sa}
(K)$. Every $A\in{\mathcal T}_{\rm sa}(H)$ has a representation
$A=A^+-A^-$ with $A^{\pm}\geq 0$ and $A^+A^-=0$. Thus $\|A\|_{\rm
Tr}=\|A^+\|_{\rm Tr}+\|A^-\|_{\rm Tr}$. As $\|\Phi(\rho)\|_{\rm
Tr}=\|\rho\|_{\rm Tr}$ for all states $\rho$, we see that
$\|\Phi(A)\|_{\rm Tr}\leq \|\Phi(A^+)\|_{\rm Tr}+\|\Phi(A^-)\|_{\rm
Tr}=\|A^+\|_{\rm Tr}+\|A^-\|_{\rm Tr}=\|A\|_{\rm Tr}$. Hence $\Phi$
is bounded and, by Corollary 2.4, $\Phi$ has the desired form.
\hfill$\Box$

\if As an end of this section, we give a characterization of affine
maps between convex sets of states,  Kadison's result.

 {\bf Corollary 2.6.} {\it Let $H$ and $K$ be    Hilbert spaces
of any dimension. Suppose that $\Phi:{\mathcal
S}(H)\rightarrow{\mathcal S} (K)$ is a bijective map. Then $\Phi$ is
affine if and only if   there exists a unitary or conjugate unitary
operator $U: H\rightarrow K$  such that $\Phi(\rho )=U\rho U^*$ for
all $\rho\in{\mathcal S}(H)$.}

{\bf Proof.} The ``if" part is obvious. For the ``only if" part,
assume that $\Phi$ is affine. As $\Phi$ is bijective, it is easily
checked that $\Phi({\mathcal Pur}(H))={\mathcal Pur}(K)$. Applying
Theorem 2.5, there exists a linear or conjugate linear operator
$U:H\rightarrow K$ with $U^*U=I_H$ such that $\Phi(\rho)=U\rho U^*$
for all$\rho$. Now $U$ is surjective since $\Phi$ preserves states
in both directions. \hfill$\Box$ \fi

\section{linear maps preserving separable pure states: bipartite systems}

Now we are ready to give a characterization of linear maps
preserving separable pure states for bipartite quantum systems.

 Write ${\mathcal
Pur}(H)\otimes{\mathcal Pur}(K)=\{P\otimes Q: P\in{\mathcal Pur}(H),
\ Q\in{\mathcal Pur}(K)\}$.

{\bf Lemma 3.1.} {\it Let $H$ and $K$ be  any Hilbert spaces. Then
the set of separable states ${\mathcal S}_{\rm sep}(H\otimes K)$ is
a convex set, whose extreme points is ${\mathcal
Pur}(H)\otimes{\mathcal Pur}(K)$.}

{\bf Proof.} Obvious. \hfill$\Box$

 Denote by ${\mathcal
T}_{\rm sep}(H\otimes K)$  the real linear space generated by
${\mathcal S}_{\rm sep}(H\otimes K)$, the set of all separable
states on $H\otimes K$; ${\mathcal F}_{\rm sep}(H\otimes K)$ the
subspace of all finite-rank operators in ${\mathcal T}_{\rm
sep}(H\otimes K)$.
 We denote by  ${\rm
Tr}_i$   the partial trace of the $i$th subsystem, that is, ${\rm
Tr}_1(\rho)={\rm Tr}_H(\rho)=({\rm Tr}\otimes I_K)(\rho)$ and ${\rm
Tr}_2(\rho)={\rm Tr}_K(\rho)=(I_H\otimes {\rm Tr})(\rho)$. Clearly,
${\rm Tr}_i$ is linear.

The following is the main result of this paper.

{\bf Theorem 3.2.} {\it Let $H$ and $K$ be two  Hilbert spaces of
any dimension. Suppose that $\Phi:{\mathcal T}_{\rm sa}(H\otimes
K)\rightarrow{\mathcal T}_{\rm sa}(H\otimes K)$ is a linear map.
Then $\Phi({\mathcal Pur}(H)\otimes {\mathcal
Pur}(K))\subseteq{\mathcal Pur}(H)\otimes {\mathcal Pur}(K)$ if and
only if one of the following holds:}

(1) {\it There exists $R_1\otimes R_2\in{\mathcal
Pur}(H)\otimes{\mathcal Pur}(K)$ such that
$$\Phi(F)={\rm Tr}(F)R_1\otimes R_2$$
for all $F\in{\mathcal F}_{\rm sep}(H\otimes K)$.}

(2) {\it There exist $R_2\in{\mathcal Pur}(K)$ and a linear or
conjugate linear isometry $U_1:H\rightarrow H$ such that
$$\Phi(F)=U_1[{\rm Tr}_2(F)]U_1^*\otimes R_2$$
 for all $F\in{\mathcal F}_{\rm sep}(H\otimes K)$.}

(3) {\it There exist
 $R_1\in{\mathcal Pur}(H)$ and a linear or conjugate linear isometry
 $U_2:K\rightarrow K$ such that
 $$\Phi(F)=R_1\otimes U_2[{\rm Tr}_1(F)]U_2^*$$
 for all $F\in{\mathcal F}_{\rm sep}(H\otimes K)$.}

(4) {\it $\dim H\geq \dim K$, there exist $R_2\in{\mathcal Pur}(K)$
and a linear or conjugate linear isometry $U_1: K\rightarrow H$ such
that
$$\Phi(F)=U_1[{\rm Tr}_1(F)]U_1^*\otimes R_2$$
for all $F\in{\mathcal F}_{\rm sep}(H\otimes K)$. }

(5) {\it $\dim H\leq \dim K$, there exist $R_1\in{\mathcal Pur}(H)$
and a linear or conjugate linear isometry $U_2:H\rightarrow K$ such
that
$$\Phi(F)=R_1\otimes U_2[{\rm Tr}_2(F)]U_2^*$$
 for all $F\in{\mathcal F}_{\rm sep}(H\otimes K)$. }

(6) {\it There exist  linear or conjugate linear isometries
$U_1:H\rightarrow H$ and $U_2:K\rightarrow K$ such that
$$\Phi(F)=(U_1\otimes U_2)F(U_1\otimes U_2)^*$$
for all $F\in{\mathcal F}_{\rm sep}(H\otimes K)$.}

(7) {\it $\dim H=\dim K$, there exist
 linear or conjugate linear isometries  $U_1: K\rightarrow H$ and
 $U_2: H\rightarrow K$ such that
 $$\Phi(F)=(U_1\otimes U_2)\theta (F)(U_1\otimes U_2)^*$$
for all $F\in{\mathcal F}_{\rm sep}(H\otimes K)$, where $\theta:
{\mathcal T}(H\otimes K)\rightarrow{\mathcal T}(K\otimes H)$ is the
swap determined by $\theta(A\otimes B)=B\otimes A$. }

(8) {\it $\dim K\leq \dim H$, there exist $R_2\in{\mathcal Pur }(K)$
and a linear map $\phi_1:{\mathcal F}_{\rm sep}(H\otimes
K)\rightarrow {\mathcal F}_{\rm sa}(H)$ such that, for each
$P\otimes Q\in{\mathcal Pur}(H)\otimes {\mathcal Pur}(K)$,
$\phi_1(P\otimes Q)=U_PQU_P^*=V_QPV_Q^*$ for some linear or
conjugate linear isometries $U_P:K\rightarrow H$, $V_Q:H\rightarrow
H$, and
$$\Phi(F)=\phi_1(F)\otimes R_2$$
for all $F\in{\mathcal F}_{\rm sep}(H\otimes K)$.}

(9) {\it $\dim H\leq \dim K$, there exist $R_1\in{\mathcal Pur }(H)$
and a linear map $\phi_2:{\mathcal F}_{\rm sep}(H\otimes
K)\rightarrow {\mathcal F}_{\rm sa}(K)$ such that, for each
$P\otimes Q\in{\mathcal Pur}(H)\otimes {\mathcal Pur}(K)$,
$\phi_2(P\otimes Q)=U_PQU_P^*=V_QPV_Q^*$ for some linear or
conjugate linear isometries $U_P:K\rightarrow K$, $V_Q:H\rightarrow
K$, and
$$\Phi(F)=R_1\otimes \phi_2(F)$$
for all $F\in{\mathcal F}_{\rm sep}(H\otimes K)$.}

We remark that, in cases (6)-(9) of Theorem 3.2, it is possible to
have one of the isometries be linear and the other isometry be
conjugate-linear.

{\bf Proof of Theorem 3.2.}   It is clear that if any one of (1)-(9)
holds, then $\Phi$ preserves separable pure states.  So we only need
to check the converse.

Assume that   $\Phi({\mathcal Pur}(H)\otimes {\mathcal
Pur}(K))\subseteq{\mathcal Pur}(H)\otimes {\mathcal Pur}(K)$.

 Define two maps
$\phi_1:({\mathcal T}_{\rm sa}(H), {\mathcal T}_{\rm
sa}(K))\rightarrow{\mathcal T}_{\rm sa}(H)$  and  $\phi_2:({\mathcal
T}_{\rm sa}(H), {\mathcal T}_{\rm sa}(K))\rightarrow{\mathcal
T}_{\rm sa}(K)$ by
$$\phi_1(A,B)={\rm Tr}_2(\Phi(A\otimes B))\quad{\rm and}\quad\phi_2(A,B)={\rm Tr}_1(\Phi(A\otimes B)).\eqno(3.1)$$

Fix a $Q\in{\mathcal Pur}(K)$. Then $\phi_1(\cdot,Q):{\mathcal
T}_{\rm sa}(H)\rightarrow{\mathcal T}_{\rm sa}(H)$ and
$\phi_2(\cdot,Q):{\mathcal T}_{\rm sa}(H)\rightarrow{\mathcal
T}_{\rm sa}(K)$ are both linear. By the assumption, we have
$$\Phi(P\otimes Q)={\rm Tr}_2(\Phi(P\otimes Q))\otimes{\rm
Tr}_1(\Phi(P\otimes Q))=\phi_1(P,Q)\otimes\phi_2(P,Q)$$ for all
$P\in{\mathcal Pur}(H)$ and all $Q\in{\mathcal Pur}(K)$. It follows
that
$$\phi_1({\mathcal Pur}(H),Q)\subseteq{\mathcal Pur}(H)\quad {\rm and}\quad \phi_2({\mathcal Pur}(H),Q)\subseteq{\mathcal Pur}(K).$$
Thus, applying Theorem 2.2 to $\phi_1(\cdot,Q)$ and
$\phi_2(\cdot,Q)$, respectively, we get that, either

(i) there is a pure state $R_{iQ} $ such that $\phi_i(A,Q)={\rm
Tr}(A)R_{iQ}$ for all $A\in{\mathcal F}_{\rm sa}(H)$, $i=1,2$, or

(ii) there is a linear or conjugate linear operator $U_{iQ}$ with
$U_{iQ}^*U_{iQ}=I_H$ such that $\phi_i(A,Q)=U_{iQ}AU_{iQ}^*$ for all
$A\in{\mathcal F}_{\rm sa}(H)$, $i=1,2$.

Similarly, for any fixed $P\in{\mathcal Pur}(H)$, considering the
maps $\phi_1(P,\cdot):{\mathcal T}_{\rm sa}(K)\rightarrow{\mathcal
T}_{\rm sa}(H)$ and $\phi_2(P,\cdot):{\mathcal T}_{\rm
sa}(K)\rightarrow{\mathcal T}_{\rm sa}(K)$, we have that, either

(i$^{\prime}$) there is a pure state $R_{iP} $ such that
$\phi_i(P,B)={\rm Tr}(B)R_{iP}$ for all $B\in{\mathcal F}_{\rm
sa}(K)$, $i=1,2$, or

(ii$^{\prime}$) there is a linear or conjugate linear operator
$U_{iP}$ with $U_{iP}^*U_{iP}=I_K$ such that
$\phi_i(P,B)=U_{iP}BU_{iP}^*$ for all $B\in{\mathcal F}_{\rm
sa}(K)$, $i=1,2$.

Observe that, for $i=1,2$, $\phi_i(\cdot,Q)$s and
$\phi_i(P,\cdot)$s are continuous on ${\mathcal F}_{\rm sa}(H)$ for
all $Q\in{\mathcal Pur}(K)$ and on ${\mathcal F}_{\rm sa}(K)$ for
all $P\in{\mathcal Pur}(H)$, respectively.

Now, we consider the map $\phi_1(\cdot,Q)$.

{\bf Claim 1.} Either $\phi_1(\cdot,Q)$ has the form (i) for all
$Q\in{\mathcal Pur}(K)$ or $\phi_1(\cdot,Q)$ has the form (ii) for
all $Q\in{\mathcal Pur}(K)$.

Fix $A_0=e_1\otimes e_1-e_2\otimes e_2\in{\mathcal F}_{\rm sa}(H)$
and define a function $F:{\mathcal Pur}(K)\rightarrow{\mathbb R}$ by
$F(Q)=\|\phi_1(A_0,Q)\|_{\rm Tr}$ for all $Q\in{\mathcal Pur}(K)$.
Note that $F(Q)=\|{\rm Tr}(A_0)R_{1Q}\|_{\rm Tr}=0$ if $\phi_1$ has
the form (i) and $F(Q)=\|U_{iQ}A_0U_{iQ}^*\|_{\rm Tr}=\|A_0\|_{\rm
Tr}= {2}$ if $\phi_1$ has the form (ii).

Take any two distinct $Q_1,Q_2\in{\mathcal Pur}(K)$. Then there
exist two linearly independent unit vectors  $x,y\in K$ such that
$Q_1=x\otimes x$ and $Q_2=y\otimes y$. For any $t\in[0,1]$, define
$$Q(t)=\frac{1}{\|x+t(y-x)\|^2}(x+t(y-x))\otimes (x+t(y-x))\in{\mathcal Pur}(K).$$
Clearly, $Q(0)=Q_1$ and $Q(1)=Q_2$. Note that, for each $t\in[0,1]$,
$\phi_1(\cdot,Q(t))$ has the form (i) or (ii). Let ${\mathcal
L}={\rm span}\{ A_0\otimes Q(t): t\in[0,1]\}$. It is clear that
${\mathcal L}$ is a finite dimensional subspace of ${\mathcal
T}_{\rm sep}(H\otimes K)$ and hence $\Phi|_{{\mathcal L}}$ is
continuous. It follows that $\phi_1(\cdot,Q)|_{{\mathcal L} }$ is
continuous and hence $t\mapsto F(Q(t))$ is a continuous map. As
$F(Q(t))$ can take only two possible distinct values, it must be a
constant. So Claim 1 holds.

Similarly, we have

{\bf Claim 1$^\prime$.} Either $\phi_2(\cdot,Q)$ has the form (i)
for all $Q\in{\mathcal Pur}(K)$ or $\phi_2(\cdot,Q)$ has the form
(ii) for all $Q\in{\mathcal Pur}(K)$.

{\bf Claim 2.} One of the following holds:

(a) For all $Q\in{\mathcal Pur}(K)$, both $\phi_1(\cdot,Q)$ and
$\phi_2(\cdot,Q)$ have the form (i).

(b) For all $Q\in{\mathcal Pur}(K)$, $\phi_1(\cdot,Q)$ has the form
(i) and $\phi_2(\cdot,Q)$ has the form (ii).

(c) For all $Q\in{\mathcal Pur}(K)$, $\phi_1(\cdot,Q)$ has the form
(ii) and $\phi_2(\cdot,Q)$ has the form (i).

We need only to check that, for all $Q\in{\mathcal Pur}(K)$,
$\phi_1(\cdot,Q)$   and $\phi_2(\cdot,Q)$ can not have the form (ii)
simultaneously. Suppose there exists some $Q_0\in{\mathcal Pur}(K)$
such that both $\phi_1(\cdot,Q_0)$ and $\phi_2(\cdot,Q_0)$ are of
the form (ii). So there exist isometric or conjugate isometric
operators $U_{1Q_0}:H\rightarrow H$ and $U_{2Q_0} : H\rightarrow K$
such that $\phi_i(A,Q_0)=U_{iQ_0}AU_{iQ_0}^*$ for all $A\in{\mathcal
F}_{\rm sa}(H)$,  $i=1,2$. Thus, we must have $\dim H\leq\dim K$ and
$$\Phi(P\otimes Q_0)=\phi_1(P,Q_0)\otimes \phi_2(P,Q_0)=U_{1Q_0}PU_{1Q_0}^*\otimes U_{2Q_0}PU_{2Q_0}^*=U(P\otimes P)U^*\eqno(3.2)$$
for all $P\in{\mathcal Pur}(H)$, where $U=U_{1Q_0}\otimes
U_{2Q_0}:H\otimes H\rightarrow H\otimes K$. Particularly, take
$P_1=e_1\otimes e_1$, $P_2=e_2\otimes e_2$,
$P_3=\frac{1}{2}(e_1\otimes e_1+e_1\otimes e_2+e_2\otimes
e_1+e_2\otimes e_2)$ and $P_4=\frac{1}{2}(e_1\otimes e_1-e_1\otimes
e_2-e_2\otimes e_1+e_2\otimes e_2)$. Then $P_1+P_2=P_3+P_4$ and so
$P_1\otimes Q_0+P_2\otimes Q_0=P_3\otimes Q_0+P_4\otimes Q_0$.
However, by a simple calculation, $P_1\otimes P_1+P_2\otimes
P_2\not=P_3\otimes P_3+P_4\otimes P_4$. Note that
$$\Phi(P_1\otimes
Q_0+P_2\otimes Q_0)=U(P_1\otimes P_1+P_2\otimes
P_2)U^*$$and$$\Phi(P_3\otimes Q_0+P_4\otimes Q_0)=U(P_3\otimes
P_3+P_4\otimes P_4)U^*.$$ It follows that $\Phi(P_1\otimes
Q_0+P_2\otimes Q_0)\not=\Phi(P_3\otimes Q_0+P_4\otimes Q_0)$, a
contradiction. So the claim is true.

Similarly, one can check that

{\bf Claim 3.} One of the following holds:

(a$^\prime$) For all $P\in{\mathcal Pur}(H)$, both $\phi_1(P,\cdot)$
and $\phi_2(P,\cdot)$ have the form (i$^\prime$).

(b$^\prime$) For all $P\in{\mathcal Pur}(H)$, $\phi_1(P,\cdot)$ has
the form (i$^\prime$) and $\phi_2(P,\cdot)$ has the form
(ii$^\prime$).

(c$^\prime$) For all $P\in{\mathcal Pur}(H)$, $\phi_1(P,\cdot)$ has
the form (ii$^\prime$) and $\phi_2(P,\cdot)$ has the form
(i$^\prime$).

{\bf Claim 4.} If (a) and (a$^\prime$) hold, then there exists
$R_1\otimes R_2\in{\mathcal Pur}(H)\otimes{\mathcal Pur}(K)$ such
that
$$\Phi(F)={\rm Tr}(F)R_1\otimes R_2$$
for all $F\in{\mathcal F}_{\rm sep}(H\otimes K)$. Hence $\Phi$ has
the form (1).

Suppose that (a) and (a$^\prime$) hold, that is, for all
$Q\in{\mathcal Pur}(K)$, we have $\phi_i(A,Q)={\rm Tr}(A)R_{iQ}$,
and, for all $P\in{\mathcal Pur}(H)$, we have  $\phi_i(P,B)={\rm
Tr}(B)R_{iP}$. Fix $P_0\in{\mathcal Pur}(H)$ and $Q_0\in{\mathcal
Pur}(K)$. Then we get
$$\phi_i(P,Q)=\phi_i(P,Q_0)=\phi_i(P_0,Q_0)=R_i.$$  Therefore,
$\Phi(P\otimes Q)=R_1\otimes R_2$ for all $P\otimes Q\in{\mathcal
Pur}(H)\otimes{\mathcal Pur}(K)$. By the linearity of $\Phi$, one
sees that Claim 4 is true.

 {\bf Claim 5.} If (a) and (b$^\prime$) hold, then $\Phi$ has the
 form (3).

 In this case, for any $P\otimes Q\in{\mathcal Pur}(H)\otimes
 {\mathcal Pur}(K)$, we have
 $$\Phi(P\otimes Q)=R_{1Q}\otimes R_{2Q}=R_{1P}\otimes
 U_{2P}QU_{2P}^*,$$
 which implies that $R_{1Q}=R_{1P}$ is independent of $P,Q$ and
 $U_{2P}QU_{2P}^*=R_{2Q}$ is independent of $P$. So there exist
 $R_{1}\in{\mathcal Pur}(H)$ and a linear or conjugate linear isometry
 $U_2$ such that $\Phi(P\otimes Q)=R_1\otimes U_2QU_2^*=R_1\otimes
 U_2[{\rm Tr}_1(P\otimes Q)]U_2^*$ for all separable pure states
 $P\otimes Q$. Now by the linearity of $\Phi$, the claim is true.

Similarly, one can show the following Claims 6-8.

{\bf Claim 6.} If (a) and (c$^\prime$) hold, then (4) holds, that
is, there exist $R_2\in{\mathcal Pur}(K)$ and a linear or conjugate
linear isometry $U_1: K\rightarrow H$ such that
$$\Phi(F)=U_1[{\rm Tr}_1(F)]U_1^*\otimes R_2$$
for all $F\in{\mathcal F}_{\rm sep}(H\otimes K)$. In this case we
must have $\dim H\geq \dim K$.

{\bf Claim 7.} If (b) and (a$^\prime$) hold, then $\Phi$ has the
form (5), that is, there exist $R_1\in{\mathcal Pur}(H)$ and a
linear or conjugate linear isometry $U_2:H\rightarrow K$ such that
$$\Phi(F)=R_1\otimes U_2[{\rm Tr}_2(F)]U_2^*$$
 for all $F\in{\mathcal F}_{\rm sep}(H\otimes K)$. In this case
 $\dim H\leq \dim K$.

{\bf Claim 8.} If (c) and (a$^\prime$) hold, then there exist
$R_2\in{\mathcal Pur}(K)$ and a linear or conjugate linear isometry
$U_1:H\rightarrow H$ such that
$$\Phi(F)=U_1[{\rm Tr}_2(F)]U_1^*\otimes R_2$$
 for all $F\in{\mathcal F}_{\rm sep}(H\otimes K)$. Hence $\Phi$
 takes the form (2).

 {\bf Claim 9.} If (b) and (c$^\prime$) hold, then $\Phi$ has the form (7). \if there exist, may not
  simultaneously,
 linear or conjugate linear isometries  $U_1: K\rightarrow H$ and
 $U_2: H\rightarrow K$ such that
 $$\Phi(F)=(U_1\otimes U_2)\theta (F)(U_1\otimes U_2)^*$$
for all $F\in{\mathcal F}_{\rm sep}(H\otimes K)$, where $\theta:
{\mathcal T}(H\otimes K)\rightarrow{\mathcal T}(K\otimes H)$ is the
swap determined by $\theta(A\otimes B)=B\otimes A$.\fi

Now suppose that (b) and (c$^\prime$) hold. Then
$$\phi_2(\cdot,Q_0)=U_{2Q_0}(\cdot)U_{2Q_0}^*\quad{\rm and}
\quad\phi_1(P_0,\cdot)=U_{1P_0}(\cdot)U_{1P_0}^*$$ with
$U_{2Q_0}^*U_{2Q_0}=I_H$ and $U_{1P_0}^*U_{1P_0}=I_K$. Moreover,
$$\phi_1(P_0,Q)={\rm Tr}(P_0)R_{1Q}={\rm Tr}(P)R_{1Q}=\phi_1(P,Q)$$
and
$$\phi_2(P,Q_0)={\rm Tr}(Q_0)R_{2P}={\rm Tr}(Q)R_{2P}=\phi_2(P,Q).$$
Thus, we obtain
$$\Phi(P\otimes Q)=\phi_1(P,Q)\otimes \phi_2(P,Q)=\phi_1(P_0,Q)\otimes \phi_2(P,Q_0)
=U_{1P_0}QU_{1P_0}^*\otimes U_{2Q_0}PU_{2Q_0}^*$$ for all
$P\in{\mathcal Pur}(H)$ and $Q\in{\mathcal Pur}(K)$. Let
$U_1=U_{1P_0}$ and $U_2= U_{2Q_0}$. Then $\Phi(P\otimes
Q)=(U_1\otimes U_2)(Q\otimes P)(U_1\otimes U_2)^* =(U_1\otimes
U_2)\theta(P\otimes Q)(U_1\otimes U_2)^*$ for all separable pure
states $P\otimes Q$. Obviously, $\dim H=\dim K$ in this case. It
follows from the linearity
  of $\Phi$ that the claim  is true.

Similarly, we have

{\bf Claim 10.} If (c) and (b$^\prime$) hold, then there exist
linear or conjugate linear isometries $U_1:H\rightarrow H$ and
$U_2:K\rightarrow K$ such that
$$\Phi(F)=(U_1\otimes U_2)F(U_1\otimes U_2)^*$$
for all $F\in{\mathcal F}_{\rm sep}(H\otimes K)$. Hence in this case
we have (6).

{\bf Claim 11.} If
 (b) and
(b$^\prime$)  hold, then $\Phi$ has the form (9).

Assume (b) and (b$^\prime$)  hold synchronously. Then for any
$P\otimes Q\in{\mathcal Pur}(H)\otimes{\mathcal Pur}(K)$, we have
$\Phi(P\otimes Q)=R_{1Q}\otimes U_{2Q}PU_{2Q}^*=R_{1P}\otimes
U_{2P}QU_{2P}^*$. It follows that there exists $R_1\in{\mathcal Pur
}(H)$ such that $R_{1Q}=R_{1P}=R_1$ and $U_{2Q}PU_{2Q}^*=
U_{2P}QU_{2P}^*$ for all $P,Q$. Thus there exists
 a linear map $\phi_2:{\mathcal F}_{\rm sep}(H\otimes K)\rightarrow
{\mathcal F}_{\rm sa}(K)$ such that, for each $P\otimes
Q\in{\mathcal Pur}(H)\otimes {\mathcal Pur}(K)$, $\phi_2(P\otimes
Q)=U_PQU_P^*=V_QPV_Q^*$ for some  linear or conjugate linear
isometries $U_P:K\rightarrow K$, $V_Q:H\rightarrow K$, and
$$\Phi(F)=R_1\otimes \phi_2(F)$$
for all $F\in{\mathcal F}_{\rm sep}(H\otimes K)$. In this case $\dim
H\leq \dim K$. So the claim is true.


Similarly,

{\bf Claim 12.}  If (c) and (c$^\prime$)    hold, then  $\Phi$ takes
the form (8).

 The proof of the
theorem is complete.\hfill$\Box$

The cases (8) and (9) of Theorem 3.2 seem not as natural as the
other forms. We do not know whether or not they may really occur. It
raises another interesting question of characterizing the real
linear maps from $ {\mathcal F}_{\rm sep}(H\otimes K)$ into $
{\mathcal F}_{\rm sa}(H)$ that send separable pure states to pure
states.

\if false The cases (8) and (9) of Theorem 3.2 seem not as natural
as the other forms, but they may occur, even for finite dimensional
cases. For example, let $H=K$. For any $P,Q\in {\mathcal Pur}(H)$,
we have ${\mathcal U}(P)={\mathcal U}(Q)$, where ${\mathcal
U}(P)=\{UPU^*: U\in{\mathcal B}(H)$ is unitary$\}$ is the unitary
orbit of $P$. Fix $P$ and a unitary $U_P$. Then there is a unitary
$U_Q$ such that $U_QPU_Q^*=U_PQU_P^*$ for any $Q$. For fixed $Q$ and
$U_Q$, there exists $U_P$ for every $P$ such that
$U_QPU_Q^*=U_PQU_P^*$ (We may even require that
$U_QPU_Q^*\not=U_{Q_1}PU_{Q_1}^*$ and
$U_PQU_P^*\not=U_{P_1}QU_{P_1}^*$ whenever $P\not=P_1$ and
$Q\not=Q_1$). For above $U_P$ and $U_Q$, let $\phi:{\mathcal T}_{\rm
sa}(H\otimes H)\rightarrow {\mathcal T}_{\rm sa}(H)$ be a
 bounded linear map determined by $\phi(P\otimes
 Q)=U_PQU_P^*=U_QPU_Q^*$. As ${\mathcal Pur}(H)\otimes{\mathcal
 Pur}(H)$ is a linearly independent set, $\phi$ is well-defined. Take any pure state $R$ and let
 $\Phi:{\mathcal T}_{\rm sa}(H\otimes H)\rightarrow{\mathcal T}_{\rm sa}(H\otimes
 H)$ be a map defined by $\Phi(A)=R\otimes \phi(A)$. Then, $\Phi$ is
 a linear map of the form (9) and preserves pure states.\fi

{\bf Corollary 3.3.} {\it Let $H$ and $K$ be two  Hilbert spaces of
any dimension. Suppose that $\Phi:{\mathcal T}_{\rm sa}(H\otimes
K)\rightarrow{\mathcal T}_{\rm sa}(H\otimes K)$ is a linear map
satisfying $\Phi({\mathcal Pur}(H)\otimes {\mathcal
Pur}(K))\subseteq{\mathcal Pur}(H)\otimes {\mathcal Pur}(K)$ and
 $\Phi({\mathcal Pur}(H)\otimes {\mathcal Pur}(K))$ contains two
elements $P^\prime_i\otimes Q^\prime_i$, $i=1,2$, with
$\{P_1^\prime, P_2^\prime\}$ and $\{Q_1^\prime,Q_2^\prime\}$
linearly independent sets. Then $\Phi$ has the form (6) or (7) in
Theorem 3.2.}

In the case of finite dimension, we get a generalization of the main
result obtained in \cite{FLPS}; there the condition $\Phi({\mathcal
Pur}(H)\otimes {\mathcal Pur}(K))={\mathcal Pur}(H)\otimes {\mathcal
Pur}(K)$ is assumed.

{\bf Corollary 3.4.} {\it Let $H$ and $K$ be two finite-dimensional
Hilbert spaces. Suppose that $\Phi:{\mathcal T}_{\rm sa}(H\otimes
K)\rightarrow{\mathcal T}_{\rm sa}(H\otimes K)$ is a linear map.
Then $\Phi({\mathcal Pur}(H)\otimes {\mathcal
Pur}(K))\subseteq{\mathcal Pur}(H)\otimes {\mathcal Pur}(K)$ if and
only if one of the statements (1)-(9) holds for all $A\in{\mathcal
T}_{\rm sep}(H\otimes K)$.}

Let $\{e_i\}_{i=1}^{\dim H}$ and $\{u_j\}_{j=1}^{\dim K}$ be
orthonormal bases of $H$ and $K$ respectively. With respect to the
product basis $\{e_i\otimes u_j\}_{i,j}$ of $H\otimes K$, the linear
map ${\bf T}\otimes {\rm id} :{\mathcal T}(H\otimes
K)\rightarrow{\mathcal T}(H\otimes K)$ determined by $A\otimes
B\mapsto A^T\otimes B$ is called the partial transpose of the first
system. The partial transpose of the second system ${\rm id}\otimes
{\bf T}$ is defined similarly. In terms of partial transposition,
one can restate Theorem 3.2 to avoid the term ``conjugate linear".
In fact, if $U$ is a conjugate isometry, then there exists an
isometry $V$ such that $UAU^*=VA^{\rm t}V^*$ for all $A$.

In the finite dimensional case, for  a linear map $\Phi$,
surjectivity and separable pure state preserving is equivalent to
preserving separable pure states in both directions, and in turn, is
equivalent to $\Phi({\mathcal Pur}(H)\otimes {\mathcal
Pur}(K))={\mathcal Pur}(H)\otimes {\mathcal Pur}(K)$. But for the
infinite dimensional case, bijectivity and separable pure state
preserving might not imply that $\Phi$ preserves  separable pure
states in both directions.

 The next result is a
characterization of linear maps preserving separable pure states in
both directions. We state it avoiding the term ``conjugate linear".

 {\bf Corollary 3.5.} {\it Let $H$ and $K$ be two
Hilbert spaces of any dimension. Suppose that $\Phi:{\mathcal
T}_{\rm sa}(H\otimes K)\rightarrow{\mathcal T}_{\rm sa}(H\otimes K)$
is a linear map. Then the following conditions are equivalent.}

(1) {\it $\Phi$ preserves separable pure states in both directions.}

(2) {\it $\Phi({\mathcal Pur}(H)\otimes {\mathcal Pur}(K))={\mathcal
Pur}(H)\otimes {\mathcal Pur}(K)$.}

(3) {\it Either }

(i) {\it There exist    unitary operators $U_1\in{\mathcal B}(H)$
and $U_2\in{\mathcal B}(K)$ such that
$$\Phi(F)=(U_1\otimes U_2)\Lambda(F)(U_1\otimes U_2)^*$$
for all $F\in{\mathcal F}_{\rm sep}(H\otimes K)$;  or}

(ii) {\it $\dim H=\dim K$, there exist
  unitary operators  $U_1\in{\mathcal B}( K, H)$ and
 $U_2\in{\mathcal B}(H,K)$ such that
 $$\Phi(F)=(U_1\otimes U_2)\Lambda(\theta (F))(U_1\otimes U_2)^*$$
for all $F\in{\mathcal F}_{\rm sep}(H\otimes K)$.\\ Here $\Lambda $
is one of the identity map, the transpose, a partial transpose with
respect to any fixed product orthonormal basis of $H\otimes K$, and
$\theta: {\mathcal T}(H\otimes K)\rightarrow{\mathcal T}(K\otimes
H)$ is the swap determined by $\theta(A\otimes B)=B\otimes A$. }

{\bf Proof.} (3)$\Rightarrow$(1)$\Rightarrow$(2) are obvious. For
(2)$\Rightarrow$(3), by Corollary 3.3, we see that $\Phi({\mathcal
Pur}(H)\otimes {\mathcal Pur}(K))={\mathcal Pur}(H)\otimes {\mathcal
Pur}(K)$ implies that the statement (6) or (7) in Theorem 3.2 holds.
Moreover, the linear or conjugate linear isometries involved are all
surjective and hence unitary or conjugate unitary. \hfill$\Box$

\if false ???{\bf Theorem 3.6.} {\it Let $H$ and $K$ be two Hilbert
spaces of any dimension. Suppose that $\Phi:{\mathcal T}_{\rm
sep}(H\otimes K)\rightarrow{\mathcal T}_{\rm sep}(H\otimes K)$ is a
linear map. Then $\Phi$  satisfies $\Phi({\mathcal S}_{\rm
sep}(H\otimes K))\subseteq{\mathcal S}_{\rm sep}(H\otimes K)$ and
$\Phi({\mathcal Pur}(H)\otimes {\mathcal Pur}(K))\subseteq{\mathcal
Pur}(H)\otimes {\mathcal Pur}(K)$ if and only if one of the
conditions (1)-(9) in Theorem 3.2 holds for all $A\in {\mathcal
T}_{\rm sep}(H\otimes K)$. }

{\bf Proof.} By Theorem 3.2 and Remark 3.5, the assumption
$\Phi({\mathcal Pur}(H)\otimes {\mathcal Pur}(K))\subseteq{\mathcal
Pur}(H)\otimes {\mathcal Pur}(K)$ implies that one of (1)-(9) holds
for all finite-rank operators in $ {\mathcal T}_{\rm sep}(H\otimes
K)$. Note that $\Phi$ is positive. Thus, for any $A,B\in {\mathcal
T}_{\rm sep}(H\otimes K)$, if $0\leq A\leq B$, then $0\leq
\Phi(A)\leq\Phi(B)$. Assume that $\rho$ is countably separable, that
is, $\rho=\sum_{i=1}^\infty t_iP_i\otimes Q_i$ for some $P_i\otimes
Q_i\in{\mathcal Pur}(H)\otimes{\mathcal Pur}(K)$ and positive
numbers $t_i$ with $\sum_{i=1}^\infty t_i=1$. Then
$\rho_n=\sum_{i=1}^n t_iP_i\otimes Q_i$ is of finite rank for each
$n$ and $0\leq \rho_n\leq \rho$. So, $\Phi(\rho_n)\leq\Phi(\rho)$
for each $n$. As  $\{\Phi(\rho_n)\}_{n=1}^\infty$ is an increasing
sequence upper bounded by $\Phi(\rho)$, $\Phi(\rho_n)$ converge to a
positive operator $\sigma\in {\mathcal T}_{\rm sep}(H\otimes K)$. It
follows that
$\sigma=\lim_{n\rightarrow\infty}\sum_{i=1}^nt_i\Phi(P_i\otimes
Q_i)=\sum_{i=1}^\infty t_i\Phi(P_i\otimes Q_i)\leq \Phi(\rho)$.
However, $\sigma$ is a state and hence we must have
$\Phi(\rho)=\sum_{i=1}^\infty t_i\Phi(P_i\otimes Q_i)$, that is, one
of (1)-(9) valid for such $\rho$.

Note that ${\mathcal F}_{\rm sep}(H\otimes K)$ is dense in
${\mathcal T}_{\rm sep}(H\otimes K)$ and the assumption
$\Phi({\mathcal S}_{\rm sep}(H\otimes K))\subseteq{\mathcal S}_{\rm
sep}(H\otimes K)$ implies that $\Phi$ is bounded. Hence the theorem
is true.\hfill$\Box$\fi

{\bf Remark 3.6.} \if Note that, if $\dim H\otimes K<\infty$, then
$\Phi({\mathcal Pur}(H)\otimes {\mathcal Pur}(K))={\mathcal
Pur}(H)\otimes {\mathcal Pur}(K)\Leftrightarrow\Phi({\mathcal
S}_{\rm sep}(H \otimes  K))={\mathcal S}_{\rm sep}(H \otimes K)$.
The implication ``$\Rightarrow$'' is not true if $\dim H\otimes
K=\infty$ because $\Phi$ may be not continuous. \fi Note that, by
the proof of Theorem 3.2, the assumption of $\Phi : {\mathcal
T}_{\rm sa}(H\otimes K)\rightarrow {\mathcal T}_{\rm sa}(H\otimes
K)$ may be replaced by the assumption of $\Phi : {\mathcal T}_{\rm
sep}(H\otimes K)\rightarrow {\mathcal T}_{\rm sep}(H\otimes K)$, and
the results of Theorem 3.2, Corollaries 3.3-3.5 remain true. This
remark is useful in some applications.

As an application of Theorem 3.2,  let us consider the separable
pure state preserving maps between states of bipartite systems.

{\bf Theorem 3.7.} {\it Let $H$ and $K$ be two Hilbert spaces of any
dimension. Suppose that $\Phi:{\mathcal S} (H\otimes
K)\rightarrow{\mathcal S}(H\otimes K)$ is an affine map. Then the
following statements are equivalent.}

(1) {\it $\Phi({\mathcal Pur}(H)\otimes {\mathcal
Pur}(K))\subseteq{\mathcal Pur}(H)\otimes {\mathcal Pur}(K)$ .}

(2) {\it The conditions (1)-(9) in Theorem 3.2 hold for all $\rho\in
{\mathcal S}_{\rm sep}(H\otimes K)$. }

{\bf Proof.} (2)$\Rightarrow$(1) is obvious.

(1)$\Rightarrow$(2). As $\Phi$ is affine, it can be extended to a
real linear map from ${\mathcal T}_{\rm sa}(H\otimes K)$ into
${\mathcal T}_{\rm sa}(H\otimes K)$ which still sends separable pure
states to separable pure states. Furthermore,  the fact
$\Phi({\mathcal S} (H\otimes K))\subseteq{\mathcal S} (H\otimes K)$
implies that $\Phi$ is continuous, and then, applying Theorem 3.2,
one sees that (2) holds.\hfill$\Box$

{\bf Corollary 3.8.} {\it Let $H$ and $K$ be two Hilbert spaces of
any dimension. Suppose that $\Phi:{\mathcal S} (H\otimes
K)\rightarrow{\mathcal S} (H\otimes K)$ is an  affine  map. Then
  the following statements are
equivalent. }

(1) {\it $\Phi$ preserves separable states in both directions.}

(2) {\it $\Phi$ preserves separable pure states in both directions.}

(3) {\it Either}

(i) {\it There exist    unitary operators $U_1\in{\mathcal B}(H)$
and $U_2\in{\mathcal B}(K)$ such that
$$\Phi(\rho)=(U_1\otimes U_2)\Lambda(\rho)(U_1\otimes U_2)^*$$
for all $\rho\in{\mathcal S}_{\rm sep}(H\otimes K)$; or}

(ii) {\it $\dim H=\dim K$, there exist
  unitary operators  $U_1\in{\mathcal B}( K, H)$ and
 $U_2\in{\mathcal B}(H,K)$ such that
 $$\Phi(\rho)=(U_1\otimes U_2)\Lambda(\theta (\rho))(U_1\otimes U_2)^*$$
for all $\rho\in{\mathcal S}_{\rm sep}(H\otimes K)$.\\ Here $\Lambda
$ is one of the identity map, the transpose, a partial transpose
with respect to any fixed product orthonormal basis of $H\otimes K$,
and $\theta: {\mathcal T}(H\otimes K)\rightarrow{\mathcal
T}(K\otimes H)$ is the swap determined by $\theta(A\otimes
B)=B\otimes A$. }

{\bf Proof.} (3)$\Rightarrow$(1) is obvious.

(1)$\Rightarrow$(2): $\Phi$ preserves separable states in both
directions implies that $\Phi|_{{\mathcal S}_{\rm sep}(H\otimes K)}:
{\mathcal S}_{\rm sep}(H\otimes K)\rightarrow{\mathcal S}_{\rm
sep}(H\otimes K)$ is a bijective affine map. Note that an affine
isomorphism between two convex sets preserves extreme points in both
directions. Hence $\Phi$ preserves separable pure states in both
directions  and (2) is true.

(2)$\Rightarrow$(3):  Assume that (2) holds; then  $\Phi({\mathcal
Pur}(H)\otimes {\mathcal Pur}(K))={\mathcal Pur}(H)\otimes {\mathcal
Pur}(K)$.
 By Theorem 3.7 and Corollary 3.5,  we see that either (i) or (ii) holds for $\Phi$ and  all
 separable states $\rho$, that is, (3) is true.  \hfill$\Box$

\section{linear maps preserving separable pure states: multipartite systems}

Results similar to that in Section 3 for bipartite cases  are valid
for multipartite cases also, but with   more complicated
expressions. The techniques of the proofs are  almost identical to
those used in the preceding part of the paper and in \cite{FLPS}. In
this section we only list those results which have relatively simple
expressions and which may have more applications. The meanings of
the notations used here are also similar to that in Section 2. For
example, ${\mathcal S}_{\rm sep}(H_1\otimes H_2\otimes \cdots\otimes
H_n)$ denotes the closed convex hull of all fully separable states
in ${\mathcal S}(H_1\otimes H_2\cdots\otimes H_n)$.

The following result corresponds to Corollary 3.3.

{\bf Theorem 4.1.} {\it Let $H_1, H_2,\ldots,H_n$ be complex Hilbert
spaces of any dimensions and let $\Phi:{\mathcal T}_{\rm
sa}(H_1\otimes H_2\otimes \cdots\otimes H_n)\rightarrow{\mathcal
T}_{\rm sa}(H_1\otimes H_2\otimes \cdots\otimes H_n)$ be a linear
map. Then $\Phi({\mathcal Pur}(H_1)\otimes\cdots\otimes{\mathcal
Pur}(H_n))\subseteq {\mathcal Pur}(H_1)\otimes\cdots\otimes{\mathcal
Pur}(H_n)$ and there are $P_1^\prime\otimes\cdots\otimes
P_n^\prime\in\Phi({\mathcal Pur}(H_1)\otimes\cdots\otimes{\mathcal
Pur}(H_n))$ and $Q_1^\prime\otimes\cdots\otimes Q_n^\prime\in
\Phi({\mathcal Pur}(H_1)\otimes\cdots\otimes{\mathcal Pur}(H_n))$
with $\{P_i^\prime,Q_i^\prime\}$ linearly independent for each
$i=1,2,\ldots, n$, if and only if  there is a permutation
$\pi:(1,\ldots, n)\mapsto (p_1,\ldots,p_n)$ of $(1,\ldots, n)$ and
linear or conjugate linear isometries  $U_j: H_{p_j}\rightarrow
H_j$, $j=1,\ldots,n$, such that }
$$\Phi(F)=(U_1\otimes\cdots\otimes U_n)\theta_\pi(F)(U_1^*\otimes\cdots\otimes
U_n^*) \eqno(4.1)
$$
{\it holds for all $F\in{\mathcal F}_{\rm
sep}(H_1\otimes\cdots\otimes H_n)$. Here $\theta_\pi:{\mathcal
T}_{\rm sa}(H_1\otimes H_2\otimes\cdots\otimes H_n)\rightarrow
{\mathcal T}_{\rm sa}(H_{p_1}\otimes H_{p_2}\otimes\cdots\otimes
H_{p_n})$ is a linear map determined by $\theta_\pi(A_1\otimes
A_2\otimes\cdots\otimes A_n)=A_{p_1}\otimes
A_{p_2}\otimes\cdots\otimes A_{p_n}$, and each $U_j$ can be
independently linear or conjugate-linear.}

It is clear that, if $\Phi$ has the form Eq.(4.1), then $\dim
H_{p_j}\leq \dim H_j$. So, if $\dim H_{p_j}> \dim H_j$ for some $j$,
then $\Phi$ cannot take the form (4.1) for the permutation $\pi:
(1,\ldots, n)\mapsto (p_1,\ldots, p_n)$.  Actually, $\dim H_j$ is
constant for all indices $j$ in a cycle of the permutation $\pi$.

The following is a special case corresponding to Theorem 3.7.

{\bf Theorem 4.2.} {\it  A  linear map $\Phi:{\mathcal T}_{\rm
sa}(H_1\otimes H_2\otimes \cdots\otimes H_n)\rightarrow{\mathcal
T}_{\rm sa}(H_1\otimes H_2\otimes \cdots\otimes H_n)$  satisfies the
conditions of Theorem 4.1 and the condition $\Phi({\mathcal S}
(H_1\otimes\cdots\otimes H_n))\subseteq {\mathcal S}
(H_1\otimes\cdots\otimes H_n)$   if and only if Eq.(4.1) holds for
all $\rho\in{\mathcal T}_{\rm sep}(H_1\otimes\cdots\otimes H_n)$.}

Particularly,

{\bf Theorem 4.3.} {\it Let $H_1, H_2,\ldots,H_n$ be complex Hilbert
spaces of any dimensions and let $\Phi:{\mathcal S} (H_1\otimes
H_2\otimes \cdots\otimes H_n)\rightarrow{\mathcal S} (H_1\otimes
H_2\otimes \cdots\otimes H_n)$ be an affine map. Then the following
statements are equivalent.}

(1) {\it $\Phi$ preserves fully separable states in both
directions.}

(2) {\it $\Phi$ preserves separable pure states in both directions.}

(3) {\it There is a permutation $\pi:(1,\ldots,
n)\mapsto(p_1,\ldots,p_n)$ of $(1,\ldots, n)$ and  unitary or
conjugate unitary operators  $U_j: H_{p_j}\rightarrow H_j$,
$j=1,\ldots,n$, such that }
$$\Phi(\rho)=(U_1\otimes\cdots\otimes U_n)\theta_\pi(\rho)(U_1^*\otimes\cdots\otimes
U_n^*) \eqno(4.2)
$$
{\it holds for all $\rho\in{\mathcal S}_{\rm
sep}(H_1\otimes\cdots\otimes H_n)$. Here, each $U_j$ can be
independently linear or conjugate-linear.}

Obviously, if $\Phi$ is of the form (4.2), then $\dim H_{p_j}=\dim
H_j$.

Theorems 4.1-4.3 may be restated in terms of linear isometric
(unitary) operators and partial transpositions as in Corollaries 3.5
and 3.8, but with more complicated expressions.

{\bf Acknowledgement.} The authors wish to give their thanks to the
referees. They read the original manuscript carefully, made up some
gaps and gave many helpful suggestions to improve the paper.

\end{document}